\documentclass[12pt, a4paper]{article}

\usepackage{amsmath}
\usepackage{amsthm}
\usepackage{amssymb}
\usepackage{mathrsfs}
\usepackage{bbm}
\usepackage{graphicx,color}

\numberwithin{equation}{section}

\newcommand{\cf}{cf. }

\newcommand{\ie}{i.e. }
\newcommand{\eg}{e.g. }

\newcommand{\iu}{\mathrm{i}}
\newcommand{\Id}{1}

\newcommand{\ideal}[1]{\mathcal{I}_{ #1 }}
\newcommand{\str}{^{*}}
\newcommand{\ep}[1]{\mathrm{e}^{#1}}
\newcommand{\hilb}{\mathcal{H}}

\newcommand{\tr}{\mathrm{tr}}

\newcommand{\T}{\mathrm{T}\,}
\newcommand{\Tstr}{\mathrm{T}^*\,}
\newcommand{\Tvec}{\vec{\mathrm{T}}}

\newcommand{\ad}[1]{\mathrm{ad}_{#1}}

\newcommand{\mean}[1]{\langle #1 \rangle}
\newcommand{\cumul}[1]{\langle\!\langle #1 \rangle\!\rangle}

\newcommand{\teop}{\hfill$\square$}

\newtheorem{thm}{Theorem}

\newtheorem{prop}[thm]{Proposition}

\newenvironment{eqwithnum}{\begin{equation}}{\end{equation}\ignorespacesafterend}
\newenvironment{eqnonum}{\begin{equation}}{\end{equation}\ignorespacesafterend}

\title{Time ordering and counting statistics}
\author{S. Bachmann ${}^{(a)}$, G.M. Graf ${}^{(a)}$, G.B. Lesovik ${}^{(a,b)}$\\
\normalsize\it
${}^{(a)}$ Theoretische Physik, ETH-H\"onggerberg, 8093 Z\"urich, Switzerland\\
\normalsize\it
${}^{(b)}$ L.D. Landau Institute for Theoretical Physics RAS, 117940 Moscow, Russia}

\begin{document}

\maketitle

\begin{abstract} 
The basic quantum mechanical relation between fluctuations of transported
charge and current correlators is discussed. It is found that, as a rule, 
the correlators are to be time-ordered in an unusual way. Instances where the
difference with the conventional ordering matters are illustrated by means of a
simple scattering model. We apply the results to resolve a discrepancy
concerning the third cumulant of charge transport across a tunnel 
junction.
\end{abstract}


\section{Introduction}
\label{intro}

Transport in mesoscopic systems has been discussed
using different approaches, based on {\it current} correlators 
\cite{L, B, BlBu}, 
either time- or in/out-ordered, or on the statistics of the transferred 
{\it charge} \cite{LL} (and, related to the latter, on the precession of a
spin coupled to current~\cite{LLL}). In this work we intend to address two
main points. First, we describe the relation between the counting statistics
and the time ordering of correlators. It turns out that the correct time
ordering differs as a rule from the conventional one, $\T$, and is given by
the Matthews' $\Tstr$-ordering~\cite{Mat}. Second, we will present a model of
energy independent scattering, where that difference matters. Though the model
is implicit in previous works~\cite{B, LLL, SaHePe}, its formalization allows to establish the equivalence between the in/out- and the $\Tstr$-ordering of currents and hence between the two approaches mentioned at the beginning.

An application of these findings is the clarification of a discrepancy between
\cite{LC} and \cite{SaHePe} concerning the third cumulant of charge transfer
through a tunnel junction. In \cite{SaHePe} the discrepancy was claimed
to be entirely due to the difference existing between unordered and 
$\T$-ordered correlators \cite{LL2, LL}. Our explanation is different, 
as we spell out momentarily.

An often chosen framework rests on states $|t,\alpha\rangle$ forming a 
basis and labelled by their time $t$ of passage across a fiducial point and 
w.r.t. a reference dynamics, 
as well as by further quantum numbers $\alpha$. To be
precise, in such a scheme $t$ is an observable which entails as its 
canonically 
conjugate operator a Hamiltonian whose spectrum similarly covers 
the real line. The description is therefore an effective one valid near the 
Fermi energy; indeed, from that perspective
the energy spectrum appears unbounded. This remark being made, 
the scattering amplitude 
from $|t_1,\alpha_1\rangle$ to $|t_2,\alpha_2\rangle$ is denoted by
$S_{\alpha_2\alpha_1}(t_2-t_1)$. Equivalence between in/out- and $\T$-ordering 
has been shown~\cite{BeSch, SaHePe} under the assumption that the scattering 
matrix satisfies
\begin{eqnonum}
S(t) = 0\,,\quad (t\le 0)\,.
\end{eqnonum}
The assumption, there called causality, should be called strict causality, as
it \eg rules out the limiting, but simple case of instantaneous,
$S(t)\propto\delta(t)$, or equivalently, energy independent
scattering. Actually, in the latter case, disagreement between the two
orderings was found~\cite{LC, Ba}. Our observation that $\Tstr$ rather than
$\T$ matters explains this difference, in that only for strictly causal
scattering the two orderings agree. Generally the difference takes the guise 
of a Schwinger term related to the infinite depth of the Fermi sea, at least 
for the lowest order cumulant where the difference matters, \ie the third. That cumulant has
been the object of experimental work \cite{Retal0, Retal}, which may be read 
as a confirmation of the $\Tstr$ ordering in the above mentioned model.

The plan of the paper is as follows. In Section~\ref{sec2} we discuss the
general relation between the generating function of the moments of the
tranferred charge and the time ordering
of current correlators. It is unrelated to scattering and not necessarily
placed within a scheme of second quantization. In Section~\ref{sec3} we
introduce a model of energy independent scattering at the level of first
quantization. We also discuss the equivalence between in/out- and 
$\Tstr$-ordering. In Section~\ref{sec4} we will promote the model to second quantization. The quantities of Section~\ref{sec2}, as well as others, can then be computed explicitly and the relation found there illustrated on this example.


\section{The $\Tstr$-ordering of current correlators}\label{sec2} 

\subsection{The result and its context}

We consider a current carrying device as symbolically illustrated in Fig.~1. The generating function of counting statistics~\cite{LL2} is
\begin{eqnonum}
\chi(\lambda, t) = \mean{\ep{\iu\lambda\Delta Q}}=
\sum_{k=0}^{\infty}\frac{(\iu\lambda)^k}{k!}\mean{(\Delta Q)^k}\,,
\end{eqnonum}
where $\mean{(\Delta Q)^k}$ is the $k$-th moment of the charge transported 
during time~$t$. Similarly, $\log \chi$ generates the cumulants 
$\cumul{(\Delta Q)^k}$.
\begin{figure}[htbp]
\begin{center}
\input{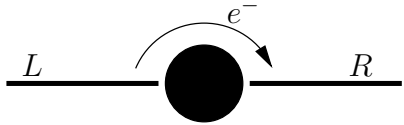} 
\end{center}
\caption{A device connecting two leads.}
\end{figure}
Several quantum mechanical expression for $\chi$ have been proposed, based
in part on different measurement protocols. The first proposal \cite{LL2}
is
\begin{eqnonum}
\chi(\lambda, t) = \mean{\exp{(\iu\lambda\smallint_0^tI(t')dt')}}
\end{eqnonum}
where $I(t)$ is the current operator at the junction and at time $t$, and 
$\mean{\cdot}$ has become the expectation in the initial quantum state of the
system. The second one
\cite{LL} reads, as recast by \cite{MA}, 
\begin{eqwithnum}
\label{LLchi}
\chi(\lambda, t) = \mean{\ep{\iu\lambda Q(t)}\ep{-\iu\lambda Q}}
=\langle \ep{\iu H t}\ep{\iu \lambda Q}\ep{-\iu H t}\ep{-\iu \lambda Q} \rangle\,,
\end{eqwithnum}
where $Q(t)=\exp(\iu H t) Q \exp(-\iu H t)$ is the charge operator on the
right of the junction and at time~$t$,
and $H$ is the Hamiltonian. The definition is appropriate to the situation 
where the initial state is an eigenstate of $Q$ and the charge is measured 
after time $t$. In fact, if the initial eigenvalue is $q$, then 
$\chi(\lambda, t) =\mean{\ep{\iu \lambda Q(t)}}\ep{-\iu\lambda q}$ just
describes the statistics of $q'-q$, where $q'$ is the (random) outcome of the
measurement of $Q(t)$. The quantity $q'-q$ is identified with the transported
charge $\Delta Q$. We mention in passing that there is a generalization
\cite{SchRa} of
this definition to the case where the assumption on the initial state does not
apply and $q$ is the outcome of the measurement of $Q(0)=Q$.

A further proposal \cite{LL3, LLL} is based on observing a spin coupled 
to current 
\begin{eqwithnum}
\label{LLLchi}
\chi(\lambda, t) = \mean{\ep{\iu H(-\lambda/2) t}\ep{-\iu H(\lambda/2) t}}
=\langle \ep{-\iu \frac{\lambda}{2} Q}\ep{\iu H t}\ep{\iu \lambda Q}\ep{-\iu H
t}\ep{-\iu \frac{\lambda}{2} Q} \rangle\,,
\end{eqwithnum}
where $H(\lambda)=\ep{\iu \lambda Q} H\ep{-\iu \lambda Q}$. This approach does
not require the initial state to be an eigenstate of $Q$, but if it does, then
it agrees with~(\ref{LLchi}).
A related proposal was put forward in \cite{LC} (with similar expressions 
found in \cite{F, BeSch}). It is given by the Keldysh time-ordered expression
\begin{eqwithnum}
\label{LC}
\chi(\lambda, t) = 
\mean{\Tvec\exp{\bigl(\iu(\lambda/2) \smallint_0^t I(t')dt'\bigr)}
\T\exp{\bigl(\iu(\lambda/2) \smallint_0^t I(t')dt'\bigr)}}\,,
\end{eqwithnum}
where $\T$ denotes the usual time-ordering and $\Tvec$ the one in the opposite
direction. For later purposes we recall that the time ordering is
supposed to occur inside the integrals once the exponential is
expanded in powers of $\lambda$. 

Eqs.~(\ref{LLLchi}) and (\ref{LC}) may 
differ in some applications, as it is the case in the very simple situation of
independent fermions passing with fixed transparency $T$ across a scatterer
biased by $V$. In fact it was found in \cite{LLL} and \cite{LC} that the 
third cumulants are, respectively, 
\begin{eqwithnum}
\label{thirdcum}
\cumul{(\Delta Q)^3}= T(1-T)(1-2T)\frac{Vt}{2\pi}\,,\qquad
\cumul{(\Delta Q)^3}= -2T^2(1-T)\frac{Vt}{2\pi}\, .
\end{eqwithnum}
In \cite{SaHePe} the first answer has been found for the second approach, 
\ie for (\ref{LC}), by 
trading the Keldysh ordering for in/out-ordering. This step in turn relies
on the strict causality of the scattering matrix, which however is not 
consistent with an energy-independent transparency.

The results of this article explain the discrepancies. 
We will consider~(\ref{LLchi}) for simplicity, though similar considerations
could be made for~(\ref{LLLchi}). As mentioned before, the two definitions agree in some
cases, including the model underlying (\ref{thirdcum}). We want to point out
that some care is to be exercised when expressing the moments in terms of
time-ordered current correlation functions. The result is 
\begin{eqwithnum}
\label{moments}
\mean{(\Delta Q)^k} = \int_0^t d^k t\,\mean{\Tstr\bigl(I(t_1)\cdots I(t_k)\bigr)}\,,
\end{eqwithnum}
where $d^k t = dt_1\cdots dt_k$, $I(t)= dQ(t) / dt$, $Q(t) = \exp(\iu H t) Q \exp(-\iu H t)$, and $\Tstr$ means that the derivative has to be taken after the time ordering:
\begin{eqwithnum}
\label{currstar}
\Tstr\bigl(I(t_1)\cdots I(t_n)\bigr) := \frac{\partial}{\partial t_n}\cdots\frac{\partial}{\partial t_1}\T\bigl( Q (t_1)\cdots Q (t_n)\bigr)\,.
\end{eqwithnum}
Hence, eq.~(\ref{LLchi}) may be summarized as 
$\chi(\lambda, t) = \mean{
\Tstr\exp{\bigl(\iu(\lambda/2) \smallint_0^t I(t')dt'\bigr)}}$; by 
(\ref{currstar}), this amounts to keeping the time ordering outside of the 
integrals of the exponential series. Similarly, 
(\ref{LLLchi}) is equivalent to (\ref{LC}) once stars are added to the time
orderings. 

It is known \cite{Mat, EpGl} and will be shown below that, as a rule, 
(\ref{currstar}) differs from $\T(I(t_1)\cdots $ $I(t_n))$ by contact 
terms
supported at coinciding times but agrees with it if $[Q, I] = 0$.


\subsection{The derivation}

In this section we show~(\ref{moments}) and obtain an expansion 
of~(\ref{currstar}) in contact terms. Any arising operators will be assumed to
be properly defined and that will be implicitely verified in later sections
in the context of applications. 

Either side of
\begin{eqwithnum}
\label{brackets}
\mean{\left(\ep{\iu H t}\ep{\iu \lambda Q}\ep{-\iu H t}\right)\ep{-\iu \lambda Q}} = \mean{\ep{\iu H t}\left(\ep{\iu \lambda Q}\ep{-\iu H t}\ep{-\iu \lambda Q}\right)}
\end{eqwithnum}
is an expression for $\chi(\lambda, t)$ and may serve as a starting point for
the derivation of~(\ref{moments}). We will follow both avenues, because the
l.h.s. leads to a simple proof, while the r.h.s. has often been considered in
the literature \cite{LLL, KiNa}.

We begin with the l.h.s., which yields
\begin{eqnonum}
\chi(\lambda, t) = \mean{\ep{\iu \lambda Q(t)}\ep{-\iu\lambda Q}} = \mean{\T \ep{\iu\lambda (Q(t)-Q)}}
\end{eqnonum}
and hence
\begin{eqnonum}
\mean{(\Delta Q)^k} = \mean{\T(Q(t)-Q)^k} = \left. \mean{\T\bigl((Q(t_1)-Q)\cdots (Q(t_k)-Q)\bigr)} \right|_{t_1 = \ldots = t_k = t}\,.
\end{eqnonum}
We note that the expression in brackets vanishes if $t_i = 0$ for some $i= 1,\ldots,n$. By repeated use of the fundamental theorem of calculus we obtain
\begin{align*}
\mean{(\Delta Q)^k} &= \int_0^{t} d t_1\,
\frac{\partial}{\partial t_1}\left.\mean{\T\bigl((Q(t_1)-Q)\cdots
(Q(t_k)-Q)\bigr)}\right|_{t_2 = \ldots = t_k = t}\\
&= \int_0^{t} d^k t\,\frac{\partial}{\partial t_k}\cdots\frac{\partial}{\partial t_1}\mean{\T\bigl((Q(t_1)-Q)\cdots (Q(t_k)-Q)\bigr)}\,.
\end{align*}
Expanding the correlator in $Q(t_i)$ and $-Q$, the resulting second term is independent of $t_i$ and does not contribute to the derivative. This proves~(\ref{moments}).

Let us now turn to the r.h.s. of~(\ref{brackets}):
\begin{eqnonum}
\chi(\lambda, t) = \mean{\ep{\iu H t}\ep{-\iu H(\lambda) t}}\,,
\end{eqnonum}
where
\begin{eqnonum}
H(\lambda) = \ep{\iu \lambda Q} H\ep{-\iu \lambda Q} = \sum_{j=0}^{\infty}\frac{(\iu\lambda)^j}{j!}\ad{Q}^j(H)
\end{eqnonum}
with multiple commutators defined by $\ad{A}^0(B) = B$, $\ad{A}^j(B) = [A, \ad{A}^{j-1}(B)]$, $(j\geq 1)$. Therefore, $\ep{\iu H t}\ep{-\iu H(\lambda) t}$ is the propagator in the interaction picture for $H$ with interaction
\begin{eqwithnum}
\label{interac}
W = H(\lambda) - H = \iu\sum_{j=1}^{\infty}\frac{(\iu\lambda)^j}{j!}\ad{Q}^{j-1}(I)\,.
\end{eqwithnum}
It is expressed by the Dyson expansion
\begin{eqnonum}
\ep{\iu H t}\ep{-\iu H(\lambda) t} = \sum_{n=0}^\infty \frac{(-\iu)^n}{n!}\int_0^t d^n t\, \T\bigl(W(t_1)\cdots W(t_n)\bigr)
\end{eqnonum}
with $W(t) = \exp(\iu H t) W \exp(-\iu H t)$. The term of order $\lambda^k$ in the expansion is obtained by picking the power $\lambda^{j_i}$ from $W(t_i)$ through~(\ref{interac}), in such a way that $\sum_{i=1}^n j_i = k$:
\begin{eqnonum}
\mean{(\Delta Q)^k} = \sum_{\substack{n, (j_1,\ldots j_n) \\ 
\sum_i j_i = k}}
\frac{k!}{n! j_1! \cdots j_n!}\int_0^{t} d^n t \,\mean{\T\bigl(\ad{Q}^{(j_1-1)}(I)(t_1)\cdots \ad{Q}^{(j_n-1)}(I)(t_n)\bigr)}\,.
\end{eqnonum}
The domain of integration is invariant under permutations of the times and the integrand under those of its factors. Thus, $n$-tuples $(j_1,\ldots,j_n)$ differing only by order may be binned together. To this end, let
\begin{eqwithnum}
\label{nj}
n_j = \sharp\{ l\:|\:j_l = j\}
\end{eqwithnum}
be the number of times the power $\lambda^j$ has been picked. Each bin then consists of
\begin{eqnonum}
\frac{n!}{n_1!\cdots n_k!}
\end{eqnonum}
tuples, since permutations among equal $j_i$'s do not generate new tuples. Thus,
\begin{eqwithnum}
\label{explmom}
\mean{(\Delta Q)^k} = \sum_{\substack{(n_1, n_2, \ldots) \\ 
\sum_j j n_j = k}}
\frac{k!}{\Bigl(\prod_{i=1}^\infty n_i!\Bigr)\Bigl(\prod_{j=1}^\infty j!^{n_j}\Bigr)}\int_0^{t} d^n t \,\mean{\T\bigl(\ad{Q}^{(j_1-1)}(I)(t_1)\cdots \ad{Q}^{(j_n-1)}(I)(t_n)\bigr)}\,,
\end{eqwithnum}
where $n = \sum_j n_j$ and the $j_l$'s satisfy~(\ref{nj}). The products are finite because of the condition $\sum_j j n_j = k$.

The derivation just given establishes the equality between (\ref{explmom}) 
and~(\ref{moments}). In the following we give an independent one by 
expanding (\ref{moments}) in contact terms.

\begin{prop}
\begin{align}
\Tstr\bigl( I (t_1)\cdots I (t_k)\bigr) &= \frac{\partial}{\partial t_k}\cdots\frac{\partial}{\partial t_1}\T\bigl( Q (t_1)\cdots Q (t_k)\bigr) \nonumber \\
& = \sum_{P\in\mathcal{P}_k} \T\Bigl( \prod_{C\in P} \ad{Q}^{n(C)-1}(I)(t_C)\delta_C \Bigr)\,,
\label{Matthews}
\end{align}
where the sum runs over all partitions $P$ of $\{1,\ldots,k\}$ into nonempty disjoint subsets $C$. By $\delta_C$ we understand a product of $\delta$-functions collapsing the times $t_i$, $(i\in C)$ to a single time $t_C = t_i$. More precisely, for $C$ consisting of $n(C)$ elements, $C = \{i_1,\ldots,i_{n(C)}\}$, we set $\delta_C = \prod_{j=1}^{n(C)-1} \delta (t_{i_j} - t_{i_{j+1}})$; in particular $\delta_C = 1$ for $n(C) = 1$.
\end{prop}

\paragraph{Remark.} The partition into single-element clusters contributes 
$\T(I(t_1)\cdots I(t_k))$.

\paragraph{Proof.} We claim the slightly more general statement for $0\leq l \leq k$:
\begin{eqwithnum}
\label{Matproof}
\frac{\partial}{\partial t_l}\cdots\frac{\partial}{\partial t_1}\T\bigl( Q (t_1)\cdots Q (t_k)\bigr) = \sum_{P\in\mathcal{P}_l} \T\Bigl( \bigl( \prod_{C\in P} \ad{Q}^{n(C)-1}(I)(t_C)\delta_C\bigr)\cdot Q(t_{l+1})\cdots Q(t_k) \Bigr)\,,
\end{eqwithnum}
where the sum now runs over $\mathcal{P}_l$ instead of $\mathcal{P}_k$. This is trivially true for $l=0$ and identical to~(\ref{Matthews}) for $l=k$. To run the induction in $l$, we note that for operators $A_i$ associated with times $t_i$, 
\begin{eqwithnum}
\label{DerT}
\frac{\partial}{\partial t_j} \T(A_1\cdots A_r) = \T(A_1\cdots \dot{A}_j \cdots A_r) + \sum^r_{i=1,i\neq j}\delta(t_j-t_i) \T([A_j, A_i]A_1\cdots\check{A}_i\cdots\check{A}_j\cdots A_r)\,,
\end{eqwithnum}
where $\,\check{}\,$ denotes omission. Using this with $j=l+1$ on~(\ref{Matproof}) we observe that $\delta(t_{l+1}-t_i) [Q(t_{l+1}), Q(t_i)] = 0$. The other commutators so generated are
\begin{eqnonum}
\delta(t_{l+1}-t_C)[Q(t_{l+1}), \ad{Q}^{n(C)-1}(I)(t_C)] = \delta(t_{l+1}-t_C)\ad{Q}^{n(C)}(I)(t_C)\,,
\end{eqnonum}
so that
\begin{align*}
\frac{\partial}{\partial t_{l+1}}&\cdots\frac{\partial}{\partial t_1}\T\bigl( Q (t_1)\cdots Q (t_k)\bigr) \\
&= \sum_{P\in\mathcal{P}_l} \T\Bigl( \bigl( \prod_{C\in P} \ad{Q}^{n(C)-1}(I)(t_C)\delta_C\bigr)\cdot I(t_{l+1})Q(t_{l+2})\cdots Q(t_k) \Bigr) \\
&\quad +\sum_{P\in\mathcal{P}_l} \sum_{C'\in P}\T\Bigl( \ad{Q}^{n(C')}(I)(t_{C'})\delta_{C'}\delta(t_{l+1}-t_C)\cdot\bigl( \prod_{\substack{C\in P \\ C\neq C'}} \ad{Q}^{n(C)-1}(I)(t_C)\delta_C\bigr) \\
&\qquad\qquad\qquad\qquad \cdot Q(t_{l+2})\cdots Q(t_k) \Bigr)\,.
\end{align*}
This agrees with~(\ref{Matproof}) for $l+1$ in place of $l$. In fact, partitions of $\{1,\ldots,l+1\}$ are distinguished by whether the cluster $\widetilde{C}$ containing $l+1$ is $\widetilde{C} = \{l+1\}$ (first line) or $\widetilde{C} = C' \cup \{l+1\}$ with $C'\in P\in\mathcal{P}_l$ (second line); in this case, $n(\widetilde{C})-1 = n(C')$.\teop

We can now verify the stated equality. After inserting~(\ref{Matthews}) in~(\ref{moments}), partitions $P\in\mathcal{P}_k$ can be binned according to the numbers $n_j$, $(j = 1,2,\ldots)$ of their clusters of size $j$ (hence $\sum_j j n_j = k$). Each bin consists of
\begin{eqnonum}
\frac{k!}{\bigl(\prod_{i=1}^\infty n_i!\bigr) \bigl(\prod_{j=1}^\infty j!^{n_j}\bigr)}
\end{eqnonum}
partitions. In fact, permutations of $k$ elements permuting or preserving clusters do not generate new partitions. Since in~(\ref{Matthews}), $\sum_C n(C) = k$ and $\sum_j n_j = n$, the equality between~(\ref{explmom}) and~(\ref{moments}) is established.


\section{The instantaneous scattering model}\label{sec3} 
\label{expl}

\subsection{The model}
\label{Model}

Transport of charge across a scatterer at zero temperature and at a small bias
ought to be determined fully by the Fermi velocity (equal to $1$ in suitable
units) and the scattering matrix at the Fermi energy. The Fermi sea is,
effectively, infinitely deep. The model we are about to
define is minimal in the sense that it is fully determined by
that matrix. The left and right reservoirs are represented by two
infinite one-dimensional leads, which are however chiral. 
The scattering between leads occurs instantaneously at a single point. The
model may be seen as describing a quantum point contact.

Let $\hilb=L^2(\mathbb{R})\oplus L^2(\mathbb{R})\cong
L^2(\mathbb{R};\mathbb{C}^2)$ be the single-particle Hilbert space of the 
model. The charge operator $Q$ is the projection onto the right lead,
\begin{eqnonum}
Q=\left(\begin{array}{cc}
0 & 0 \\ 0 & 1
\end{array}\right)\,,
\end{eqnonum}
the second quantization of which will later represent the number of 
particles there.
In absence of interaction, states $\psi\in \hilb$ evolve in time according to
$\psi(x-t)$. The corresponding Hamiltonian $H_0$ is linear in the momentum:
\begin{eqwithnum}
H_0 = \left(\begin{array}{cc}
p & 0 \\ 0 & p
\end{array}\right)\,,
\end{eqwithnum}
where $p = -\iu d/dx$ and the derivative is taken in the sense of
distributions. Scattering between leads is determined by the matrix
\begin{eqwithnum}
S = \left(\begin{array}{cc}
\mathfrak{r} & \mathfrak{t'} \\ \mathfrak{t} & \mathfrak{r'}
\end{array}\right)\,.
\end{eqwithnum}
We will specify the resulting Hamiltonian \cite{FG} at the end of this 
section. First however, we define the unitary group generated by it, because it is simpler and it is all which is needed in the rest of the paper. It is
\begin{eqwithnum}
\label{U}
(U(t)\psi)(x) = \left\{ 
\begin{array}{ll} \psi(x-t) + (S-1)\theta(0<x<t)\psi(x-t)\,, & (t>0)\,, \\
\psi(x-t) + (S\str-1)\theta(t<x<0)\psi(x-t)\,, & (t<0)\,,
\end{array} \right.
\end{eqwithnum}
and is motivated by the idea that the part of the freely evolved wave function
$\psi(x-t)$ which, say for times $t>0$, has crossed the scatterer at $x=0$
between times $0$ and $t$ gets replaced by $S\psi(x-t)$. One readily verifies that $U(t)$ is a strongly continuous $1$-parameter group.

Letting $U(t)$ act on the charge operator, we get
\begin{eqwithnum}
\label{Qt}
Q(t) = U(t)\str Q U(t) = \left\{
\begin{array}{ll} 
Q (\theta (-x-t) + \theta (x)) + S\str Q S \theta (x+t)\theta(-x) \,,&(t>0)\,, \\
Q (\theta (-x) + \theta (x+t)) + SQS\str \theta (-x-t)\theta (x) \,,&(t<0)\,,
\end{array} \right.
\end{eqwithnum}
or equivalently
\begin{eqwithnum}
\label{QtBis}
Q(t) = \bigl(Q \theta (-x-t) + S\str Q S \theta (x+t)\bigr)\theta(-x) + \bigl(SQS\str  \theta (-x-t) + Q \theta (x+t)\bigr)\theta(x)\,.
\end{eqwithnum}
It yields the current
\begin{eqwithnum}
\label{Current}
I(t) = \frac{d Q(t)}{dt} = \bigl((S\str Q S - Q)\theta(-x) + (Q-SQS\str)\theta(x)\bigr)\delta(x+t)\,,
\end{eqwithnum}
which is well-defined as a distribution in $t$.

We now come to the description of the generator $H$, a self-adjoint operator by Stone's theorem. To this end, let
\begin{eqnonum}
\mathcal{D}_+ = \{\psi_+\in L^2(\mathbb{R}_+;\mathbb{C}^2)\mid\psi'_+\in L^2(\mathbb{R}_+;\mathbb{C}^2)\}\,,
\end{eqnonum}
where the derivative $\psi'_+$ is that of a distribution on 
$\mathbb{R}_+ = (0, \infty)$, \ie away from the origin. Any 
function $\psi\in\mathcal{D}_+$ is continuous up to the boundary $x=0$, and let $\psi_+(0)$ be its boundary value. Any function in $L^2(\mathbb{R}_+;\mathbb{C}^2)$ may be seen as a distribution on all of $\mathbb{R}$. In this sense, the derivative of $\psi_+\in\mathcal{D}_+$ is
\begin{eqwithnum}
\label{psi+der}
\frac{d}{dx}\psi_+ = \psi'_+(x)+\psi_+(0)\delta(x)\,.
\end{eqwithnum}
We may similarly define $\psi_-\in\mathcal{D}_-$ based on $\mathbb{R}_- = (-\infty, 0)$. Then,
\begin{eqwithnum}
\label{psi-der}
\frac{d}{dx}\psi_- = \psi'_-(x) - \psi_-(0)\delta(x)\,.
\end{eqwithnum}
Given $\psi\in L^2(\mathbb{R};\mathbb{C}^2)$, let $\psi_\pm$ be their restrictions to the half-lines $\mathbb{R}_\pm$.
\begin{prop}
The generator $H$ of $U(t)$ has domain
\begin{eqnonum}
D(H) = \{\psi\in\hilb\mid\psi_\pm\in\mathcal{D}_\pm,\,\psi_+(0) = S\psi_-(0)\}
\end{eqnonum}
and is given by
\begin{eqnonum}
(H\psi)(x)= -\iu(\psi_+'(x)+\psi_-'(x))\,.
\end{eqnonum}
\end{prop}
We recall that $D(H)$ consists of those states $\psi\in\hilb$ for which the limit $\lim_{t\to 0} t^{-1}(U(t)-1)\psi = -\iu H \psi$ exists, thereby defining $H\psi$.

\paragraph{Proof.} Let first $t>0$. Then
\begin{eqnonum}
(U(t)\psi)(x) = \psi_+(x-t) + \psi_-(x-t) + (S-1)\theta(x)\psi_-(x-t)\,.
\end{eqnonum}
Using~(\ref{psi+der}, \ref{psi-der}), we find
\begin{align}
\label{Hfirst}
\frac{U(t)-1}{t}\psi\stackrel{t\downarrow 0}{\longrightarrow} &-\frac{d}{dx}\psi_+ -\frac{d}{dx}\psi_- + (S-1)\psi_-(0)\delta \\
&=-(\psi'_+ + \psi'_-)+(S\psi_-(0)-\psi_+(0))\delta
\label{Hbis}
\end{align}
in the sense of distributions if $\psi_\pm\in\mathcal{D}_\pm$. Actually, this last condition is implied by the requirement $\psi\in D(H)$. In fact, by using $t^{-1}(\phi, (U(t)-1)\psi) = t^{-1}((U(t)-1)\phi, \psi)$ on a test function $\phi$ supported away from $x=0$, we obtain in the limit $-\iu(\phi, H\psi) = (\phi_+'+\phi_-', \psi)$ and hence $\psi_\pm'\in L^2(\mathbb{R}_\pm, \mathbb{C}^2)$. At that point $\psi\in D(H)$ further implies $S\psi_-(0) = \psi_+(0)$. Moreover, the convergence~(\ref{Hbis}) is attained in $\hilb$. The same conclusion in reached for $t<0$, whence (\ref{Hbis}) is $-\iu H\psi$.\teop

More casually, the Hamiltonian can also be written as
\begin{eqwithnum}
\label{Hamiltonian}
H = H_0 + \iu(S-1)\cdot\delta_- = H_0 - \iu(S\str-1)\cdot\delta_+ = H_0 + 2\iu\frac{S-1}{S+1}\cdot\delta\,,
\end{eqwithnum}
with quadratic forms $\delta_\pm$ on $\mathcal{D}_+\cap\mathcal{D}_-$ defined as $(\delta_\pm \psi)(x) := \delta(x) \psi(0\pm)$
and $\delta = (\delta_+ + \delta_-)/2$. The first expression in~(\ref{Hamiltonian}) corresponds to~(\ref{Hfirst}) and the remaining ones follow from the boundary condition. Eq.~(\ref{Hamiltonian}) is a variant of one found in~\cite{AK}.

We note in passing that~(\ref{Current}) may alternatively be understood as a quadratic form on $D(H)$ for fixed $t$. The expression is unambiguous even at $t=0$, because $(S\str Q S - Q)\delta_-$ and $(Q-SQS\str)\delta_+$ agree as a result of the boundary condition.


\subsection{Comparison of time orderings}

The $\Tstr$-ordering of currents is identical to the in/out-ordering, $\widetilde{\T}$, whose definition \cite{BeSch} we shall recall momentarily. For the time being, we establish this fact at the level of first quantization, but we shall show in Section 4 that it persists under second quantization. The current~(\ref{Current}) may be split, $I(t) = I_+(t) + I_-(t)$, into outgoing and incoming parts,
\begin{eqwithnum}
\label{InOutCurr}
\begin{array}{lcl}
I_+(t) & = & \bigl(S\str Q S \theta(-x) + Q\theta(x)\bigr)\delta(x+t)\,, \\
I_-(t) & = & -\bigl(Q \theta(-x) + SQS\str\theta(x)\bigr)\delta(x+t)\,.
\end{array}
\end{eqwithnum}
In particular, $[I_\pm(t), I_\pm(s)] = 0$. 

\begin{prop}
\label{LmaTstrTtilde}
We have
\begin{eqwithnum}
\label{TstrTtilde}
\Tstr(I(t_1)\cdots I(t_k)) = \widetilde{\T}(I(t_1)\cdots I(t_k))\,,
\end{eqwithnum}
where the ordering $\widetilde{\T}$ places any $I_-(t)$ to the right of any $I_+(t')$, regardless of $t\gtrless t'$ (the order of currents of the same type is irrelevant), and extends by linearity to $I(t) = I_+(t) + I_-(t)$.
\end{prop}

Before giving the proof, let us make an observation. The proof of~(\ref{Matthews}) was based on
\begin{eqwithnum}
\label{GoodCom}
\frac{\partial}{\partial s}\T(I(t)Q(s)) = \T(I(t)I(s)) + \delta(t-s)[Q(s), I(t)]\,,
\end{eqwithnum}
which is a sensible equation in the context of our model. The formal analogue
\begin{eqwithnum}
\label{BadCom}
\frac{\partial}{\partial s}\T(I_+(t)Q(s)) = \T(I_+(t)I(s)) + \delta(t-s)[Q(s), I_+(t)]
\end{eqwithnum}
is meaningless on its r.h.s., and should not be used. These claims are based on the commutators
\begin{eqwithnum}
\begin{array}{lcl}
[Q(s), I_+(t)] & = & \theta(t-s)\delta(x+t)\bigl([Q, S\str Q S]\theta(-x) + [SQS\str, Q]\theta(x)\bigr)\,, \\
\left[Q(s), I_-(t)\right] & = & \theta(s-t)\delta(x+t)\bigl([Q, S\str Q S]\theta(-x) + [SQS\str, Q]\theta(x)\bigr)\,, \\
\left[I(s), I_+(t) \right] & = & -\delta(t-s)\delta(x+t)\bigl([Q, S\str Q S]\theta(-x) + [SQS\str, Q]\theta(x)\bigr)\,,
\end{array}
\label{SomeCom}
\end{eqwithnum}
which can be obtained from~(\ref{QtBis}, \ref{InOutCurr}). The first commutator is discontinuous at $t-s = 0$, hence its multiplication with $\delta(t-s)$ in~(\ref{BadCom}) is ambiguous. Also the first term there is seen to exhibit an ambiguity proportional to $\delta(t-s)$. By contrast, the sum of the first two commutators,
\begin{eqwithnum}
\label{QIComm}
[Q(s), I(t)] = \delta(x+t)\bigl([Q, S\str Q S]\theta(-x) + [SQS\str, Q]\theta(x)\bigr)\,,
\end{eqwithnum}
is independent of $s$, whence the r.h.s. of~(\ref{GoodCom}) is well-defined.

\paragraph{Proof.} The inductively stable generalization of~(\ref{TstrTtilde}) is
\begin{eqwithnum}
\label{IndTstrTtilde}
\frac{\partial}{\partial t_l}\cdots\frac{\partial}{\partial t_1}\T\bigl( Q (t_1)\cdots Q (t_k)\bigr) = \widetilde{\T}(I(t_1)\cdots I(t_l);Q(t_{l+1})\cdots Q(t_{k}))\,,\qquad(0\leq l \leq k)\,,
\end{eqwithnum}
where, on the r.h.s., the $Q$'s are time-ordered and the $I_-$ (resp. $I_+$) are placed to their right (resp. left). Upon differentiating the r.h.s. w.r.t. $t_{l+1}$, eq.~(\ref{DerT}) generates equal time commutators only among $Q$'s, which vanish. Thus,
\begin{eqnonum}
\frac{\partial}{\partial t_{l+1}}\widetilde{\T}(I(t_1)\cdots I(t_l);Q(t_{l+1})Q(t_{l+2})\cdots Q(t_{k})) = \widetilde{\T}(I(t_1)\cdots I(t_l);I(t_{l+1})Q(t_{l+2})\cdots Q(t_{k}))\,,
\end{eqnonum}
indicating that $I(t_{l+1})$ is still subject to time ordering relatively to $Q(t_{l+2})\cdots Q(t_{k})$. After splitting it, $I(t_{l+1}) = I_+(t_{l+1})+I_-(t_{l+1})$, the incoming part $I_-(t_{l+1})$ commutes with the earlier $Q$'s on its right by~(\ref{SomeCom}); similarly, $I_+(t_{l+1})$ with those on its left. The result is~(\ref{IndTstrTtilde}) with $l+1$ instead of $l$.\teop


\section{Second quantization}\label{sec4} 
\label{SecQuant}

\subsection{Binomial statistics}

For fermionic many-body systems consisting of independent particles, the generating function~(\ref{LLchi}) has been computed by Levitov and Lesovik~\cite{LL} as
\begin{eqwithnum}
\label{LLBin}
\chi(\lambda, t) = \bigl((1-T)+\ep{\iu\lambda}T\bigr)^{Vt/2\pi}\cdot (1+o(t))\,,\qquad(t\to\infty)\,,
\end{eqwithnum}
where $T = |\mathfrak{t}|^2$ is the transmission probability and $V$ is the bias across the scatterer. The result describes a binomial distribution with $N = tV / 2\pi$ attempts. In particular, it yields the third cumulant
\begin{eqwithnum}
\label{thirdCumul}
\cumul{\Delta Q^3}_\rho = \frac{Vt}{2\pi} T(1-T)(1-2T)\,.
\end{eqwithnum}
Before presenting a derivation of~(\ref{LLBin}) among others~\cite{LL, Has}, let us make a digression on second quantization. In vague terms, the second quantization $\widehat{A}$ of a single-particle operator $A$ is the sum of its contributions from all particles $i$, $\widehat{A} = \sum_i(A_i-\mean{A_i}_\rho)$, though with expectation value subtracted. The notion can be formalized for infinitely many particles in the frame of the GNS space of the multi-particle state $\rho$, \ie a Hilbert space containing $\rho$ and its local perturbations as vectors. We shall do for $\rho$ being a quasi-free fermionic state, determined by a single-particle density matrix of the same name, $1\geq \rho = \rho\str \geq 0$. In the special case of a pure many-particle state, \ie for $\rho=\rho^2$, $A$ admits a second quantization $\widehat{A}$ acting on the GNS space iff
\begin{eqwithnum}
\label{ImplCond}
[A, \rho]\in\ideal{2}\,,
\end{eqwithnum}
where $\ideal{2}$ denotes the Hilbert-Schmidt operators. The exponentiated version of this criterion is: A unitary operator $U$ admits an implementation $\widehat{U}$ on the GNS space iff $[U, \rho]\in\ideal{2}$. The notation is slightly abusive as the construction differs from $\widehat{A}$: The propagator $U = \exp(-\iu H t)$ is promoted to $\widehat{U} = \exp(-\iu \widehat{H} t)$. With these preliminaries, the generating function~(\ref{LLchi}) reads
\begin{eqwithnum}
\label{chi2Quant}
\chi(\lambda) = \left\langle\widehat{U}\str\ep{\iu\lambda\widehat{Q}}\widehat{U}\ep{-\iu\lambda\widehat{Q}}\right\rangle_\rho\,,
\end{eqwithnum}
where, in line with the hypothesis made there, we assume
\begin{eqwithnum}
\label{RhoQComm}
[\rho, Q] = 0\,.
\end{eqwithnum}
The generating function~(\ref{chi2Quant}) can be expressed in terms of the first quantized operators through the Levitov-Lesovik (infinite) determinant~\cite{LL}. In~\cite{ABGK} a particle-hole symmetric variant of the determinant was given a mathematical foundation on the basis of the GNS space. We recall the result, in the case of a pure state. Set $A_U:=U\str A U$.
\begin{prop}
Let
\begin{eqwithnum}
\label{URhoComm1}
[U, \rho]\in\ideal{1}\,,
\end{eqwithnum}
where $\ideal{1}\ (\subset\ideal{2})$ denotes the trace class operators. Then
\begin{eqnonum}
\chi(\lambda) = \det\left(1 + (\ep{-\iu\lambda}-1)Q_U\rho_U\rho' + (\ep{\iu\lambda}-1)Q_U\rho_U'\rho\right)\,,
\end{eqnonum}
where $\rho' = 1-\rho$ is the density matrix of holes. The determinant is Fredholm.
\end{prop}

In the context of~(\ref{LLBin}) and of the instantaneous scattering model~(\ref{Hamiltonian}), the initial density matrix is the projection
\begin{eqnonum}
\rho = \left(
\begin{array}{cc}
\rho_L & 0 \\ 0 & \rho_R
\end{array}
\right)
\end{eqnonum}
with $\rho_i = \theta(\mu_i -p)$, $(i=L, R)$ representing two Fermi seas
biased by $V=\mu_L-\mu_R$. While assumption~(\ref{RhoQComm}) holds true, (\ref{URhoComm1}) unfortunately does not. The failure is however of minor importance. It can be traced back to the discontinuities of $(U(t)\psi)(x)$ at $x=0$ and $x=t$ introduced by $U$, as seen in~(\ref{U}). Technically, the hypothesis could be met by making the scattering matrix time-dependent with $S(\tau) = S$ for most of the relevant time interval $\tau\in(0,t)$, but $=1$ near its ends (see~\cite{ABGK} for a similar model). The particular choice of rounding would affect the determinant only to sub-leading order in $t$ (\cf a $\log t$ contribution to $\chi(\lambda)$ in~\cite{LLL}). We neglect such contributions from the result, which in view of $\theta A_U\theta = \theta S\str A S \theta$ for $\theta = \theta(-t<x<0)$ becomes
\begin{eqwithnum}
\label{detWH}
\chi(\lambda, t) = \det{}_{\hilb(t)}\left(1+(\ep{-\iu\lambda}-1)S\str Q\rho S \rho' + (\ep{\iu\lambda}-1)S\str Q\rho' S \rho\right)\,,
\end{eqwithnum}
where $\hilb(t) = L^2([-t, 0];\mathbb{C}^2)$. The determinant is now of a form considered by Kac~\cite{Kac} and Akhiezer~\cite{Akh}, extended to the matrix case~\cite{BotSilb}: Consider a translation invariant operator $A$ on $L^2(\mathbb{R};\mathbb{C}^n)$ with Fourier multiplier $A(k)$. For its truncation to $[-t, 0]$, one has
\begin{eqwithnum}
\log \det{}_{\hilb(t)} (1+A) \longrightarrow\: \frac{t}{2\pi}\int_{-\infty}^\infty dk\,\log \det(1+A(k)) + o(t)\,,\qquad (t\to\infty)\,,
\label{BinStat}
\end{eqwithnum}
where the determinant on the r.h.s is that of an $n\times n$ matrix. We apply the result to~(\ref{detWH}) with $n=2$ and
\begin{eqnonum}
S\str Q \rho S \rho' = \rho_R\rho_L' \left(
\begin{array}{cc}
T & 0 \\ \overline{\mathfrak{r}}'\mathfrak{t} & 0
\end{array}
\right)\,,
\end{eqnonum}
as well as with $\rho$ and $\rho'$ interchanged. For $\mu_L>\mu_R$ we have $\rho_R\rho_L' = 0$ and $\rho_R'\rho_L = \theta(\mu_R<p<\mu_L)$. We so obtain
\begin{eqnonum}
\log\chi(\lambda, t) = \frac{Vt}{2\pi}\log(1+(\ep{\iu\lambda}-1)T) + o(t)\,,\qquad(t\to\infty)\,,
\end{eqnonum}
as announced.


\subsection{The third cumulant}

In this section we apply the expansion in contact terms, eq.~(\ref{Matthews}),
in order to compute the third cumulant of the transported charge. 
The currents will have to be understood as second quantized operators, which 
we emphasize by adding a hat. Eq.~(\ref{moments}) reads
\begin{eqwithnum}\label{momentsbis}
\cumul{\Delta Q^3}_\rho = \int_0^t d^3t\,\cumul{\Tstr\bigl(\widehat{I}(t_1)\widehat{I}(t_2)\widehat{I}(t_3)\bigr)}_\rho
\end{eqwithnum}
and the result to be re-derived is~(\ref{thirdCumul}). It agrees with the binomial statistics as derived again in the previous section and the result~\cite{SaHePe} based on the $\widetilde{\T}$ time ordering; on the other hand, it should be contrasted with
\begin{eqwithnum}
\label{noContact}
\int_0^t d^3t\,\cumul{\T\bigl(\widehat{I}(t_1)\widehat{I}(t_2)\widehat{I}(t_3)\bigr)}_\rho = -\frac{Vt}{2\pi} 2T^2(1-T)
\end{eqwithnum}
which is the result obtained for $\cumul{\Delta Q^3}_\rho$ in other
schemes~\cite{LL2, LC}. The difference between (\ref{momentsbis}) and 
(\ref{noContact}) is accounted for by contact terms. Indeed, in eq.~(\ref{Matthews}) with $k=3$, the partitions $P$ are $(1)(2)(3)$, $(12)(3)$, $(23)(1)$, $(31)(2)$, $(123)$, which yields
\begin{eqwithnum}
\cumul{\Delta{Q}^3}_\rho = \int_0^{t} d^3t\,\cumul{\T \widehat{I}_1\widehat{I}_2\widehat{I}_3}_\rho +3\int_0^{t} d^2t\,\cumul{\T \widehat{I}_1[\widehat{Q}_2, \widehat{I}_2]}_\rho +\int_0^{t} dt_1\,\cumul{[\widehat{Q}_1, [\widehat{Q}_1, \widehat{I}_1]]}_\rho\,,
\label{3Cumul}
\end{eqwithnum}
where $\widehat{Q}_i = \widehat{Q}(t_i)$, $\widehat{I}_i = \widehat{I}(t_i)$. Besides of~(\ref{noContact}), we will show that the two further integrals are $0$, resp. $(Vt/2\pi)\cdot T(1-T)$, resulting in
\begin{eqnonum}
\cumul{\Delta Q^3}_\rho = \frac{Vt}{2\pi} (-2T^2(1-T) + 0 + T(1-T)) = \frac{Vt}{2\pi} T(1-T)(1-2T)\,.
\end{eqnonum}
It should also be remarked in passing that $[Q,\rho]=0$, whence the
hypothesis about the initial state, discussed after eq.~(\ref{LLchi}) and
underlying eq.~(\ref{momentsbis}), is satisfied. 

Before going into the proper proof of these claims, let us recall the following elementary rules~\cite{Lund}, which reduce computations of correlators and commutators of second quantized operators to the level of first quantization:
\begin{align}
\mean{\widehat{A}}_\rho &= 0\,,\nonumber \\
\label{Meanprod}
\cumul{\widehat{A} \widehat{B}}_\rho &= \mean{\widehat{A} \widehat{B}}_\rho = \tr(\rho A \rho' B\rho )\,, \\
\cumul{\widehat{A} \widehat{B} \widehat{C}}_\rho &= \tr(\rho A \rho' B\rho'C\rho) - \tr(\rho A \rho' C \rho B \rho)\,,\nonumber
\end{align}
with $\rho' = 1-\rho$. The r.h.s. of (\ref{Meanprod}) may be cast as 
$\tr(\rho A \rho' B \rho)=\tr([\rho, A] \rho' [B ,\rho])$, which is finite 
for $A, B$ enjoying~(\ref{ImplCond}). Eq.~(\ref{Meanprod}) implies
\begin{eqwithnum}
[\widehat{A}, \widehat{B}] =\widehat{[A,B]} + \bigl(\tr(\rho A \rho' B \rho) -
\tr(\rho'A \rho B \rho')\bigr)\cdot\Id
\equiv \widehat{[A,B]} + S(A,B)\cdot\Id\,,
\label{Schwinger}
\end{eqwithnum}
where the last term is known as a Schwinger term. It implies
\begin{eqwithnum}
\hat{A}(t)=\widehat{A(t)}+\iu\int_0^tdt'\,S(H,A(t'))\Id\,.
\label{Schwinger1}
\end{eqwithnum}
In our case, the current $\widehat{I}(t) = d\widehat{Q}(t) / dt$ is a
distribution in $t$. In order to give rise to an operator on
which the above may be applied, a test function is required, which will however
remain implicit. This being said, the contact terms in (\ref{3Cumul}) are, 
by (\ref{Schwinger}, \ref{Schwinger1}),
$\cumul{\T \widehat{I}_1[\widehat{Q}_2, \widehat{I}_2]}_\rho=
\cumul{\T \widehat{I}_1\widehat{[Q_2,I_2]}}_\rho$ and
$\cumul{[\widehat{Q}_1, [\widehat{Q}_1, \widehat{I}_1]]}_\rho=
\cumul{\widehat{[Q_1, [Q_1, I_1]]}}_\rho$. They are of importance since 
the commutator
\begin{eqwithnum}
\label{commIQ}
[Q(t), I(t)] = \bigl([Q, S\str Q S]\theta(-x) + [SQS\str, Q]\theta(x)\bigr) \delta(x+t)\,,
\end{eqwithnum}
resulting from~(\ref{QIComm}), does not vanish.

The computation of the integrals~(\ref{3Cumul}) will make repeated use of the following expressions. The translation invariant density matrices $\rho$ and $\rho'$ have integral kernels
\begin{align*}
\rho(x,y) &= \frac{1}{2\pi\iu}\frac{1}{x-y-\iu 0}\cdot D(x-y)\,, \\
\rho'(x,y) &= \frac{1}{2\pi\iu}\frac{1}{y-x-\iu 0}\cdot D(x-y)
\end{align*}
with
\begin{eqnonum}
D(z) = \left(\begin{array}{ll}
\ep{\iu \mu_L z} & 0 \\ 0 & \ep{\iu \mu_R z}
\end{array}\right)\,.
\end{eqnonum}
Moreover,
\begin{eqnonum}
S\str Q S = \left(
\begin{array}{ll}
T & \mathfrak{r}'\overline{\mathfrak{t}} \\ \overline{\mathfrak{r}}'\mathfrak{t} & 1-T
\end{array}
\right)\,,\qquad[Q,S\str Q S] = \left(
\begin{array}{ll}
0 & -\mathfrak{r}'\overline{\mathfrak{t}} \\ \overline{\mathfrak{r}}'\mathfrak{t} & 0
\end{array}
\right)\,.
\end{eqnonum}
We compute the first integrand~(\ref{3Cumul}) by temporarily dropping the time ordering.
\begin{align}
\label{III}
\cumul{\widehat{I}_1\widehat{I}_2\widehat{I}_3}_\rho &= \tr(I_1 \rho' I_2 \rho' I_3 \rho) - \tr(I_1 \rho' I_3 \rho I_2 \rho)\\
&= \frac{1}{(2\pi\iu)^3}\frac{\tr\bigl(A(t_2-t_1)A(t_3-t_2)A(t_1-t_3)\bigr)-\tr\bigl(A(t_3-t_1)A(t_2-t_3)A(t_1-t_2)\bigr)}{(t_1-t_2-\iu 0)(t_2-t_3-\iu 0)(t_1-t_3-\iu 0)}\label{IIIbis}
\end{align}
with
\begin{eqwithnum}
\label{A}
A(t_i-t_j) = (S\str Q S -Q) D(t_i-t_j) = \left(\begin{array}{ll}
T & \mathfrak{r}'\overline{\mathfrak{t}} \\ \overline{\mathfrak{r}}'\mathfrak{t} & -T
\end{array}\right)D(t_i-t_j)\,,
\end{eqwithnum}
where we used (\ref{Current}). That results in
\begin{eqwithnum}
\label{IIIter}
\cumul{\widehat{I}_1\widehat{I}_2\widehat{I}_3}_\rho = \frac{-4 T^2(1-T)}{(2\pi)^3}\cdot\frac{\sin V(t_1-t_2)+\sin V(t_2-t_3)+\sin V(t_3-t_1)}{(t_1-t_2-\iu 0)(t_2-t_3-\iu 0)(t_1-t_3-\iu 0)}\,,
\end{eqwithnum}
where $V = \mu_L-\mu_R$. In fact, the first trace equals
\begin{eqnonum}
2\iu T^2(1-T)\sin V(t_3-t_1)+(\text{cyclic})
\end{eqnonum}
by the following argument. It may be expressed as $\sum_{i,j,k = L,R}A_{ij}A_{jk}A_{ki}$. The word $ijk=LLL$ involves a single diagonal entry of~(\ref{A}) with overall compensating phases. Its contribution, $T^3$, cancels an opposite contribution from the similar term $RRR$. For $LRR$, we find
\begin{eqnonum}
\mathfrak{r}'\overline{\mathfrak{t}}\,\ep{\iu\mu_R(t_2-t_1)}\cdot(-T)\ep{\iu\mu_R(t_3-t_2)} \cdot\overline{\mathfrak{r}}'\mathfrak{t}\,\ep{\iu\mu_L(t_1-t_3)}  = -T^2(1-T)\ep{-\iu V(t_3-t_1)}\,,
\end{eqnonum}
which can be combined with $T^2(1-T)\ep{\iu V(t_3-t_1)}$ from $RLL$. The
remaining words provide the cyclic permutations. To conclude, it suffices to
note that the second trace is obtained by exchanging $2\leftrightarrow 3$,
which induces a change of sign in all the exponents.

We then note that~(\ref{IIIter}) is locally integrable and we drop the regularizations. Eq.~(\ref{noContact}) then follows from
\begin{eqwithnum}
\label{AsInt}
\int_0^{t} d^3t\,\frac{\sin V(t_1-t_2)+\sin V(t_2-t_3)+\sin V(t_3-t_1)}{(t_1-t_2)(t_2-t_3)(t_1-t_3)} = 2\pi^2 Vt + o(t)\,,\quad(t\to\infty)\,.
\end{eqwithnum}
Actually, the time ordering should have been reinstated, but since the
expression is permutation symmetric that is superfluous. Eq.~(\ref{AsInt}) can
be derived by reinstating the regularizations $-\iu 0$; then the integral can
be broken into three terms, the first two of which vanish in the limit of
large $t$. In fact, the first term has poles at $t_3 = t_2-\iu 0$, $t_1-\iu 0$
which do not pinch the real axis. The same applies to the second and to the $t_1$-integration. Using $\int dt_2 (t_1-t_2-\iu 0)^{-1}\cdot(t_2-t_3-\iu 0)^{-1} = 2\pi\iu(t_1-t_3-\iu 0)^{-1}$, one is left with
\begin{eqwithnum}
2\pi\iu\int_0^t dt_1dt_3\,\frac{\sin V(t_3-t_1)}{(t_1-t_3-\iu0)^2} 
=t\cdot2\pi\iu\int dx \,\frac{\sin Vx}{(x+\iu0)^2}+ o(t) = 2\pi^2 V t + o(t)\,;
\label{AsIntBis}
\end{eqwithnum}
in fact, the odd part of $(x+\iu 0)^{-2}$ is 
$\bigl((x+\iu 0)^{-2}-(x-\iu 0)^{-2}\bigr)/2
=\pi\iu\delta'(x)$.

We next consider the middle integral in~(\ref{3Cumul}). We find
\begin{align*}
\cumul{\widehat{I}_1[\widehat{Q}_2, \widehat{I}_2]}_\rho = \tr\left(I_1\rho'[Q_2, I_2]\rho\right) &= \frac{1}{(2\pi\iu)^2}\frac{1}{(t_1-t_2-\iu 0)^2}\cdot\tr\bigl(A(t_2-t_1)[Q, S\str Q S] D(t_1-t_2)\bigr) \nonumber\\
&=\frac{1}{(2\pi\iu)^2}\frac{1}{(t_1-t_2-\iu 0)^2}\cdot\sin V(t_1-t_2)\,,
\end{align*}
and the same result for $\cumul{[\widehat{Q}_2, \widehat{I}_2]\widehat{I}_1}_\rho$ except for $+\iu 0$ instead of $-\iu 0$. Thus, $\cumul{\T\widehat{I}_1[\widehat{Q}_2, \widehat{I}_2]}_\rho$ is odd in $t_1\leftrightarrow t_2$, and the integral vanishes.

Finally the last integrand in~(\ref{3Cumul}), being the expectation value of a commutator, reduces to its Schwinger term
\begin{eqnonum}
\cumul{[\widehat{Q}_1, [\widehat{Q}_1, \widehat{I}_1]]}_\rho = \tr(\rho Q_1 \rho' [Q_1, I_1] \rho) - \tr(\rho' Q_1 \rho[Q_1, I_1]\rho')\,.
\end{eqnonum}
Using (\ref{commIQ}, \ref{Qt}) it equals
\begin{multline*}
\frac{1}{(2\pi\iu)^2}\int_{-t_1}^0dy\,\left(\frac{1}{(y+t_1+\iu 0)^2}-\frac{1}{(y+t_1-\iu 0)^2}\right)\tr(S\str Q S D(t_1+y) [Q, S\str Q S] D(-t_1-y)) \\
= \frac{-2\iu}{(2\pi\iu)^2}T(1-T)\int_{-t_1}^0dy\,\sin V(y+t_1)\left(\frac{1}{(y+t_1+\iu 0)^2}-\frac{1}{(y+t_1-\iu 0)^2}\right)\,,
\end{multline*}
where we noted that additional contributions from $y<-t_1$ and $y>0$ vanish because of $\tr(Q  D(t_1+y) [Q, S\str Q S] D(-t_1-y))=0$. After changing the integration variable to $x=y+t_1$, both factors are odd, whence the integral equals 
\begin{eqnonum}
\frac{1}{2}\int_{-t_1}^{t_1} dx\,\sin Vx
\left(\frac{1}{(x+\iu 0)^2}-\frac{1}{(x-\iu 0)^2}\right)
= -\pi\iu V+o(1)\,,
\end{eqnonum}
as in (\ref{AsIntBis}). We conclude that
\begin{eqwithnum}
\int_0^{t} dt_1\,\cumul{[\widehat{Q}_1, [\widehat{Q}_1, \widehat{I}_1]]}_\rho = T(1-T)\frac{Vt}{2\pi}\,,
\end{eqwithnum}
as claimed.


\subsection{A strictly causal scattering process}

We may trade the instantaneous scattering process in use in the previous
section with a strictly causal one, all while remaining within the model of
Section~\ref{sec3}. That is achieved by simply forfeiting a piece of length 
$l>0$
of the leads in favor of the scatterer. So reinterpreted, the scattering
process lasts $2l$ and the contact terms should disappear. Nevertheless the
same result for the third cumulant will be obtained, though only the first 
terms on the r.h.s. of~(\ref{3Cumul}) will contribute.

In physical terms we place the detector a distance $l>0$ away from the
scatterer; in mathematical terms we replace $Q$ by its regularization $Q_l = Q
\theta(|x|\geq l)$ and hence $I$ by $I_l = \iu[H, Q_l] = Q[\delta(x-l) -
\delta(x+l)]$. It ought to be noted that $[Q_l, \rho]\neq 0$, so that the
physical appropriateness of~(\ref{LLchi}), and hence of~(\ref{momentsbis}), could be questioned. Nevertheless, we may just view $l>0$ as a regulator before taking the limit $l\to 0$. However, even this is troublesome, at least in this form, because $[Q_l, \rho]\notin \ideal{2}$ makes $\widehat{Q}_l$ undefined. That can be remedied by smoothing the step function $\theta(|x|\geq l)$ on a length $<l$. With these preliminaries taken, the contact terms in~(\ref{3Cumul}) vanish as expected. This follows from $[\widehat{Q}_l, \widehat{I}_l]\propto 1$ and in turn from $[Q_l, I_l] = 0$, while the corresponding Schwinger term no longer vanishes. On the other hand, the current correlator $\cumul{\T \widehat{I}_{l1} \widehat{I}_{l2} \widehat{I}_{l3}}$ in~(\ref{3Cumul}) has a well-defined limit when the smoothing is removed. Omitting the time ordering,
it is for times $t_i>l$~\cite{Ba} 
\begin{align}\label{correl}
\cumul{\widehat{I}_{l1}\widehat{I}_{l2}\widehat{I}_{l3}}_\rho
&=T(1-T)(1-2T) \cdot \frac{2}{(2\pi)^3} \cdot \frac{\sin V(t_1-t_2)+\sin V(t_2-t_3)+\sin V(t_3-t_1)}{(t_1-t_2)(t_2-t_3)(t_1-t_3)}\nonumber \\
&\quad-T(1-T) \cdot \frac{2}{(2\pi)^3}\left( \frac{\sin V(t_1-t_2)}{(t_1-t_2)(t_2-t_3-2l-\iu 0)(t_1-t_3-2l-\iu0)} +
\right. \nonumber\\
&\qquad\qquad\qquad\qquad\left. \frac{\sin V(t_2-t_3)}{(t_1-t_2+2l-\iu 0)(t_2-t_3)(t_1-t_3+2l-\iu0)} + \right.\nonumber  \\
&\qquad\qquad\qquad \qquad\left. \frac{\sin V(t_3-t_1)}{(t_1-t_2-2l-\iu0)(t_2-t_3+2l-\iu 0)(t_1-t_3)}\right)\\
&= \mathrm{I} + \mathrm{II}\,.\nonumber 
\end{align}
We remark that the expression reduces to (\ref{IIIter}) in the limit $l\to 0$;
the integral of the latter is (\ref{noContact}) and should be distinguished
from the limit of the integral. We add that time-unordered correlators,
suitably symmetrized, can be measured by means of 
detectors discussed in \cite{Ba2}.

The derivation of (\ref{correl}) begins as in (\ref{III}), but with currents 
\begin{eqwithnum}
\label{Ilt}
I(t) = S\str Q S\delta(x-l+t) - Q \delta(x+l+t)\equiv I_+(t) + I_-(t)
\end{eqwithnum}
split into outgoing and incoming parts. The terms involving only outgoing
parts yield $\mathrm{I}$. Indeed, their contribution has the structure 
(\ref{IIIbis}, \ref{A}), but with $S\str Q S -Q$ replaced by $S\str Q S$.
In the description used there in reference to the first trace (\ref{IIIbis}), 
the words $LLL$ and $RRR$ contribute $T^3$ and 
$(1-T)^3$, respectively; $LRR$ and $RLL$ contribute 
$T(1-T)^2\ep{-\iu V(t_3-t_1)}$ and $T^2(1-T)\ep{\iu V(t_3-t_1)}$, with the 
remaining words providing cyclic permutations thereof. Again, the second trace
is obtained by flipping signs in the exponents, and the difference of 
the two is 
\begin{eqnonum}
-2\iu T(1-T)^2\sin V(t_3-t_1) + 2\iu T^2(1-T)\sin V(t_3-t_1) 
= -2\iu T(1-T)(1-2T) \sin V(t_3-t_1)\,,
\end{eqnonum}
plus cyclic permutations. Terms involving precisely one
incoming currents yield $\mathrm{II}$. Finally, terms containing more than one
incoming current vanish.

Each term of $\mathrm{II}$ exhibits two singularities not cancelled by the numerator. The terms differ however in regard as to whether the singularities may be attained within the region $t_1\geq t_2\geq t_3$: both in the first term, neither in the second, and $t_1 = t_2+2l$ but not $t_2 = t_3-2l$ in the last one. Within said region the regularization of the unaccessible singularities may be changed from $-\iu 0$ to $+\iu 0$. The so modified expression $\mathrm{II}'$ is permutation symmetric, like $\mathrm{I}$. This proves
\begin{eqnonum}
\cumul{\T \widehat{I}_{l1}\widehat{I}_{l2}\widehat{I}_{l3}} = \mathrm{I} + \mathrm{II}'
\end{eqnonum}
for all $t_1, t_2, t_3\ge l$. We can now compute its integral over $0\leq t_i\leq t$, $(i=1,2,3)$, for large $t$. The contribution from $\mathrm{I}$ is inferred from~(\ref{AsInt}). The three terms of $\mathrm{II}'$ contain singularities in the variables $t_3, t_1$, respectively $t_2$, which do not pinch the real axis. Hence $\mathrm{II}'$ does not contribute and the result is
\begin{eqnonum}
\cumul{(\Delta Q_l)^3} = \int_0^t d^3 t\,\cumul{\T \widehat{I}_{l1}\widehat{I}_{l2}\widehat{I}_{l3}} = \frac{Vt}{2\pi}T(1-T)(1-2T)+o(t)\,.
\end{eqnonum}
%


\subsection{Comparison of time orderings (continued)}

The purpose of this subsection is to show that eq.~(\ref{TstrTtilde}) persists upon replacing the currents by their second quantized counterparts:
\begin{eqnonum}
\Tstr\bigl(\widehat{I}(t_1)\cdots \widehat{I}(t_k)\bigr) = \widetilde{\T}\bigl(\widehat{I}(t_1)\cdots \widehat{I}(t_k)\bigr)\,.
\end{eqnonum}
Inspection of the proof of Proposition~\ref{LmaTstrTtilde} shows that we need to establish $[\widehat{Q}(s), \widehat{I}_+(t)] = 0$, ($t<s$) and a similar property for $\widehat{I}_-(t)$. By time covariance it suffices to prove the claim 
for $s=0$, where $Q(0)=Q$. Since by~(\ref{SomeCom}) we have
\begin{eqwithnum}
[Q(s) , I_+(t)] = 0,\quad (t<s)\,,
\label{last}
\end{eqwithnum}
at the level of first quantization, we have to make sure that the property is not destroyed by Schwinger terms:
\begin{eqnonum}
\tr\bigl(\rho Q(s) \rho' I_+(t) \rho \bigr) - \tr\bigl(\rho' Q(s) \rho I_+(t) \rho' \bigr) = 0\,,
\end{eqnonum}
which for $s=0$ indeed follows from $\rho Q \rho'=0$. Alternatively, one 
can verify that the Schwinger term does not change under a common translation
of $t$ and $s$, \ie under conjugation of $Q(s)$ and $I_+(t)$ by the 
propagator $U(\tau)$. Schwinger terms satisfy the Hochschild condition
$S(AB,C)+S(BC,A)+S(CA,B)=0$ and hence
\begin{eqnonum}
S([A,B],C)+S([B,C],A)+S([C,A],B)=0\,.
\end{eqnonum}
As a result of (\ref{last}) we have 
\begin{eqnonum}
-\iu\frac{d}{d\tau}S(Q(s+\tau),I_+(t+\tau))
=S([H,Q(s+\tau)],I_+(t+\tau))+S(Q(s+\tau),[H,I_+(t+\tau)])=0\,.
\end{eqnonum}

\noindent
{\bf Acknowledgements.} We thank A. Lebedev, M. Reznikov, M. Suslov for
valuable discussions. Support from the Russian Foundation for Basic Research 
(grant No. 08-02-00767-a) and from the Center for Theoretical Studies at ETH
is gratefully acknowledged.


\end{document}